\documentclass[12pt,a4paper,english,aps,preprint]{revtex4}
\usepackage[english]{babel}
\usepackage{graphicx,color}
\usepackage{latexsym}
\usepackage{amsmath}
\usepackage{amsfonts}
\usepackage{amssymb}
\usepackage[latin1]{inputenc}
\usepackage{hyperref}
\usepackage{simplewick,verbatim}

\begin{document}

\title{Thermofield Dynamics for Twisted Poincaré-Invariant Field Theories:
Wick Theorem and S-matrix}
\author{Marcelo Leineker}
\email{marcelo@uft.edu.br}
\affiliation{{Instituto de F\'isica,
Universidade de Bras\'ilia, Bras\'ilia, DF, 70910.900,
Brazil}\\
{\ N\'ucleo de F\'isica, Universidade Federal do Tocantins, 77001-090,
Palmas, TO, Brazil}}
\author{Amilcar R. Queiroz}
\email{amilcarq@unb.br} \affiliation{{Instituto de F\'\i sica and
International Center for Condensed Matter Physics, Universidade de
Bras\'\i lia, 70910-900, Bras\'\i lia, DF, Brazil}}
\author{Ademir E. Santana}
\email{asantana@unb.br} \affiliation{{Instituto de F\'\i sica and
International Center for Condensed Matter Physics, Universidade de
Bras\'\i lia, 70910-900, Bras\'\i lia, DF, Brazil}}
\author{Chrystian de Assis Siqueira}
\email{chrystian@uft.edu.br}
\affiliation{N\'ucleo de F\'isica, Universidade Federal do Tocantins, 77404-970, Gurupi,
TO, Brazil}

\begin{abstract}
	Poincar\'e invariant quantum field theories can be formulated on
non-commutative planes if the statistics of fields is twisted. This is
equivalent to state that the co-product on the Poincar\'e group is suitably
twisted. In the present work we present a twisted Poincar\'e invariant quantum
field theory at finite temperature. For that we use the formalism of
Thermofield Dynamics (TFD). This TFD formalism is extend to incorporate
interacting fields. This is a non trivial step, since the separation in
positive and negative frequency terms is no longer valid in TFD. In particular,
we prove the validity of Wick's theorem for twisted scalar quantum field at
finite temperature.
\end{abstract}

\begin{comment}
\begin{abstract}
Twisted Poincar\'e-invariant quantum scalar field theory deals
with a quantum scalar field in Groenenwold-Moyal plane, so that
the scalar field is the usual one, but its statistics is twisted.
Therefore it is suitable to describe many-particles systems that
might show a twisted (generalized) boson  nature. Such twisted
behavior of boson  fields may show up in very high energy
phenomena. Temperature effects certainly play a key role in the
investigation of such very high energy phenomena. Thus, it is
relevant the finite temperature characterization of such twisted
quantum field theory. In this work, we use the formalism of
Thermofield Dynamics (TFD) to twisted quantum scalar field
theories. In particular, we show how to introduce interaction
among these fields. This is not a trivial step, since the
separation in positive and negative frequencies terms for the
scalar fields is no longer valid for in TFD. We then prove the
validity of Wick theorem for these finite temperature twisted
scalar quantum fields.
\end{abstract}
\end{comment}

\maketitle

%\tableofcontents

%\maketitle

\section{Introduction}

Field theory at finite temperature is of paramount importance in
describing several aspects of many-body systems and  particle
physics,  involving for instance spontaneous symmetry breaking or
the restoration of symmetry, as in super-conductivity and in high
energy physics~\cite{umezawa93,Khanna09}. For this reason, there is a
strong interest in the extension of thermal quantum field theories
to the realm of fields associated with aspects of nature such as
non-commutativity\cite{Douglas2001,Szabo2003a}.

A central ingredient of a non-commutative quantum field theory on
Groenenwold-Moyal plane ~\cite{Connes1994,Seiberg1999a} is
the mixing of ultraviolet and infrared divergences (UV-IR mixing)
in the perturbation theory~\cite{Minwalla2000}. This UV-IR mixing, being a
signature
of non-commutativity due to the appearance of non-planar diagrams
in perturbative expansions, is relevant for applications of
non-commutativity in condensed matter systems, including the
finite temperature cases, as in the quantum hall
effect~\cite{Susskind2001}. In addition, temperature effects are
important as a way for testing non-commutativity in space
coordinates, considering the limit of high temperature~\cite{Arcioni}.
These results have  motivated different studies involving
temperature aspects
\cite{Fischler2000,Strelchenko:2007xh,Akofor2009a,Balachandran2008,Barosi2008,
Das2008,Basu:2010qm,Costa2010},
using the imaginary formalism \cite{Matsubara1955a} or thermofield
dynamics (TFD) \cite{umezawa93,Khanna09}, i.e., a real time formalism that in
equilibrium is equivalent to the Keldish-Schwinger formalism
\cite{Schwinger1961,Keldysh1964a}.

The imaginary time formalism takes the Boltzmann factor, $\ \exp (-\beta H)$%
, where $\beta $ is the inverse of temperature $T$ ($\beta =1/T$, with
$\kappa_B\equiv 1$),
under a Wick rotation in the time evolution, such that time $t$ is
mapped into a imaginary time: $t\rightarrow t=i\tau ,$ with $0\leq
\tau \leq \beta .$ The theory satisfies the KMS (Kubo, Martin and
Schwinger) boundary condition, and then the propagator is written
as a Fourier series in the imaginary time, by using the Matsubara
frequencies: $\omega _{n}=2\pi n/\beta $ ($\omega _{n}=\pi (2n+1)/\beta $) for
bosons (fermions), corresponding to the period $\beta =T^{-1},$
with $T$ being the temperature. As as consequence, in the momentum
space, the original theory is reduced to a 3-dimensional Euclidean
formalism, involving an infinite summation of the Matsubara
frequencies. However imaginary time formalism can be applied just to systems in
equilibrium. This problem does not exist in real time formalism.

On the other hand, TFD is fully structured on
algebraic methods using the notion of $\mathbb{C}^{\ast
}$-algebras~\cite{Ojima1981a, Santana1999, Santana2000} and synthesized
in two ingredients: a doubling in the Hilbert space (a consequence of the
commutants of the von Neumann algebra used in the definition of
the $\mathbb{C}^{\ast }$-algebra) and a Bogoliubov transformation, i.e., a
rotation involving the two distinct types of linear spaces,
leading to the thermalization of the theory. A consequence of such
an apparatus is that there is no imaginary time in the theory and
the propagator is written in two pieces: one describes the $T=0$
theory, while the other gives rise to temperature effects. This
aspect is useful when effects of temperature are competing with
other parameters of a theory as is the case of
non-commutative approach. This fact has been explored in the
context of non-commutativity, in particular for twisted Poincar\'e-
invariant theories \cite{Balachandran2010c}.

In quantum field theory, a crucial property is the statistics of particles. That
is encapsulated in the problem of how the
states of the system behaves under the interchange of two
particles. This many particle property has an entangled relation
with the spin nature of the identical particles that forms the
system. On the other hand, the spin nature of these particles is
related to the representation of the Poincar\'e group. For
space-time dimension strictly greater than 3, there are only two
possible classes of spin, half-integer and integer, leading to two
classes of statistics, fermions and bosons, respectively. This is
the celebrated spin-statistics theorem (See for example the book
\cite{Weinberg-vol1}.).

Recently, it was shown that if space-time is noncommutative, e.g.
$\left[ x^{\mu },x^{\nu }\right] =i\theta ^{\mu \nu}$ with $\theta ^{\mu \nu
}=-\theta ^{\nu \mu}$ and $\theta ^{0i}\neq 0,$, then it is possible to
construct a field theory invariant under Poincar\'e group, but the
spin-statistics theorem is
modified~\cite{Balachandran2006c,Balachandran2006d,Balachandran2007e,
Akofor2007,Balachandran2007e,Balachandran2007f,Balachandran:2008gr,
Balachandran2008b,Akofor2008a,Balachandran2009b,Balachandran2009c,
Balachandran2009e, Balachandran2010,Balachandran2010e} . This type of theory
has been called twisted-Poincar\'e field theories and lies in the
fact that in order to have Poincar\'e invariance, one has to twist
the action of the group in the tensor states, i.e. in the many
particles states. That is possible by twisting the co-product
action of the Poincar\'e group into the many particles Hilbert
space and such a procedure leads to non-trivial results when it is
applied, for instance, to some process in quantum chromodynamics.
Despite the fact that temperature effects play an import role in
non-commutative approaches, only some preliminary aspects of
thermal phenomena have been explored in the context of twisted
Poincar\'e-invariant theories \cite{Akofor2009a,Balachandran2010c}.

The main goal in the present work is to develop a thermal perturbative
quantum-field approach of a many-particle twisted Poincar\'e system. In order to
accomplish such a result, we use TFD, exploring several aspects of its
algebraic structure. We first prove the Wick theorem in TFD under general
basis. One of the advantages of such demonstration is that it allows to
implement interactions in TFD formalism. This result is up to our best
knowledge not available in the literature. It is therefore an important result
for TFD in its own right and its applications. We then extend this formalism to
twisted Poincar\'{e} invariant quantum field theories at finite temperature.

The present work is organized in the following manner. In the next
section, we present an outline of TFD, emphasizing some aspects to
be explored later. In Section 3, we demonstrate the validity of
Wick theorem in the TFD formalism. In Section 4, after
recollecting some facts about twisted Poincar\'e quantum field
theories, we demonstrate the validity of the thermal Wick theorem
in this non-commutative context. In Section 5, concluding remarks
are presented.

\section{The thermofield dynamics propagator}

In this section, we present an outline of thermofield dynamics, in
order to fix the notation and to emphasize aspects to be explored
later, as the thermal Green's function. We consider the scalar
field, only, for simplicity. Let $\mathcal{O}$ be a physical
observable. Its thermal average can be written as
\begin{equation*}
\langle \mathcal{O}\rangle _{\beta }=\frac{1}{Z(\beta )}~\mathrm{Tr}(%
\mathcal{O}\rho (\beta )),
\end{equation*}%
where
\begin{equation*}
Z(\beta )=\sum_{n}e^{-\beta E_{n}},
\end{equation*}%
is the canonical partition function, with $E_{n}$ being the $n$-th
eigenvalue of a Hamiltonian $H$, and
\begin{equation*}
\rho (\beta )=\frac{1}{Z(\beta )}\sum_{n}e^{-\beta E_{n}}|n\rangle \langle
n|,
\end{equation*}%
is the corresponding density matrix.

Now, let $\mathcal{O}=A(t^{\prime })B(t)$, where $A(t^{\prime })$ and $B(t)$
are any two operators in instants $t^{\prime }$ and $t$, respectively, given
in the Heisenberg picture by
\begin{equation*}
A(t^{\prime })=e^{it^{\prime }H}~A~e^{-it^{\prime }H}~\hspace{0.3cm}%
~B(t)=e^{itH}~B~e^{-itH}.
\end{equation*}%
The KMS (Kubo-Martin-Schwinger) condition reads
\begin{equation*}
\langle A(t^{\prime })B(t)\rangle _{\beta }=\langle B(t-i\beta )A(t^{\prime
})\rangle _{\beta }=\langle B(t)A(t^{\prime }-i\beta )\rangle _{\beta }.
\end{equation*}%
In the Euclidean quantum field theory at finite temperature, i.e. $%
t\rightarrow i\tau $, the KMS condition implies that correlation functions
are periodic, with period $\beta $. 

TFD is introduced by a doubling in the degrees of freedom of thermofield theory
\cite{Khanna09}:
\begin{align*}
(A_{i}A_{j})\widetilde{}& =\widetilde{A}_{i}\widetilde{A}_{j}, \\
(cA_{i}+A_{j})\widetilde{}& =c^{\ast }\widetilde{A}_{i}+\widetilde{A}_{j}, \\
(A_{i}^{\dagger })\widetilde{}& =(\widetilde{A}_{i})^{\dagger }, \\
(\widetilde{A}_{i})^{\widetilde{}}& =-\xi A_{i},
\end{align*}%
with $\xi =-1$ for bosons and $\xi =+1$ for fermions; such that the physical
variables are described by non-tilde operators. The tilde variables are
defined in the commutant of the von Neumann algebra of  quantum operators
and are associated with generators of the modular group given by $\widehat{A}%
=A-\widetilde{A}$. With these elements, reducible representations
of Lie-groups can be studied, in particular, kinematical
symmetries as the Poincar\'e group.  The other basic ingredient of
TFD is a Bogoliubov transformation, $U(\alpha )$, introducing a
rotation in the tilde and non-tilde variables, such that thermal
effects emerge from a condensate state. The rotation parameter
$\alpha $ is associated with temperature, and this procedure leads
to the usual statistical thermal average.

Consider $\phi (x)$ a real scalar field. The TFD prescription leads us then
to the doubling:  $\phi (x)\rightarrow \phi (x)\otimes \widetilde{\phi }(x)$%
. These operator fields are expanded in modes as
\begin{eqnarray}
\phi (x) &=&\int d\mu (p)\left( c(p)e^{ipx}+c(p)^{\dagger
}e^{-ipx}\right)
\label{eq:field1} \\
\widetilde{\phi }(x) &=&\int d\mu (p)~\left( \widetilde{c}(p)e^{-ipx}+%
\widetilde{c}(p)^{\dagger }e^{ipx}\right) ,  \label{eq:field2}
\end{eqnarray}%
where $d\mu (p)$ represents an invariant measure in momentum space (for the
case of continuous field it can be written as usual like $d\mu
(p)=d^{3}p/((2\pi )^{3}2p_{0})$ and $c(p)^{\dagger },\widetilde{c}%
(p)^{\dagger },c(p),\widetilde{c}(p)$ are the creation and annihilation
operators that act on the Fock space of many-particle states.

We write Eqs.~(\ref{eq:field1}) and (\ref{eq:field2}) in a compact matrix
form
\begin{eqnarray*}
\phi (x)_{I} &=&\int d\mu (p)~\left( c(p)_{J}e_{p}+c(p)_{J}^{\dagger
}e_{-p}\right) \\
&=&\phi ^{+}(x)_{J}+\phi ^{-}(x)_{J},
\end{eqnarray*}%
where $e_{p}=e^{ipx}$, and
\begin{eqnarray*}
\phi ^{+}(x)_{1} &=&\int d\mu (p)c(p)e_{p},\qquad \phi ^{-}(x)_{2}=\int d\mu
(p)\tilde{c}(p)e_{-p} \\
\phi ^{-}(x)_{1} &=&\int d\mu (p)c^{\dagger }(p)e_{-p},\quad \phi
^{+}(x)_{2}=\int d\mu (p)\tilde{c}(p)^{\dagger }e_{p}.
\end{eqnarray*}
Here the superscripts $+$ and $-$ denotes the positive and negatives frequencies, respectively, frequencies of the Fourier representation of the fields.

The Bogoliubov transformation is introduced by the canonical operator
\begin{equation*}
U_{\beta }=e^{-iG_{\beta }},
\end{equation*}%
where
\begin{equation*}
G_{\beta }=~i\int d\mu (p)~f_{\beta }(p)~\left( c(p)^{\dagger }\widetilde{c}%
(p)^{\dagger }-c(p)\widetilde{c}(p)\right) ,
\end{equation*}%
with $f(\beta )$ being defined intrinsically by the equation
\begin{equation}
\tanh (f_{\beta }(p))=e^{-\frac{\beta E_{p}}{2}},  \label{angle1}
\end{equation}%
with $E_{p}$ being the dispersion relation for bosons. The origin
of this relation lies in the Bose-Einstein distribution. Applying
the Bogoliubov transformation in the usual Poincar\'e-invariant
vacuum state $|0\rangle $, one obtains a thermal Poincar\'e-
invariant vacuum state
\begin{equation}
|0;\beta \rangle \equiv U_{\beta }|0\rangle .  \label{eq:thermal-vacuum1}
\end{equation}%
This vacuum state is annihilated by the thermal annihilation operator $%
c_{\beta }(p)\equiv U_{\beta }~c(p)~U_{\beta }^{-1}$, i.e.
\begin{equation}
~\hspace{0.3cm}~~c_{\beta }(p)|0;\beta \rangle =0.  \label{eq:thermal-annih1}
\end{equation}%
Thermal particle states are created by thermal creation operator
\begin{equation}
c_{\beta }(p)^{\dagger }\equiv U_{\beta }~c(p)^{\dagger }~U_{\beta }^{-1}
\label{eq:thermal-creation1}
\end{equation}%
acting on the thermal vacuum, Eq.~(\ref{eq:thermal-vacuum1}).

We write now the doubled operators in terms of thermal operators. For that
we use the Bogoliubov transformation, such that
\begin{equation*}
c(p)_{J}=(U_{\beta }^{-1})_{JK}~c_{\beta }(p)_{K},
\end{equation*}%
with
\begin{equation*}
c(p)_{J}=\left(
\begin{array}{c}
c(p) \\
\tilde{c}(p)^{\dagger }%
\end{array}%
\right) \qquad c_{\beta}(p)_{J}=\left(
\begin{array}{c}
c_{\beta }(p) \\
\tilde{c}_{\beta }(p)^{\dagger }%
\end{array}%
\right) ,
\end{equation*}%
and
\begin{equation*}
(U_{\beta })_{JK}=\left(
\begin{array}{cc}
u ~%
\hspace{0.2cm}~& v \\
v ~%
\hspace{0.2cm}~& u%
\end{array}%
\right)
\end{equation*}%
is the matrix representation of Bogoliubov transformation, where
\begin{equation*}
u\equiv u(f_{\beta }(p))=\frac{e^{\beta \epsilon }}{\sqrt{e^{\beta \epsilon }-1}},~%
\hspace{0.3cm}~v\equiv v(f_{\beta }(p))=\frac{1}{\sqrt{e^{\beta \epsilon }-1}}.
\end{equation*}%
Then we write $\phi(x)_{A}$, the representation of the doubled
field in terms of thermal operators, as
\begin{equation}
\phi (x)_{J}=\phi _{\beta }^{+}(x)_{J}+\phi _{\beta }^{-}(x)_{J}.\label{eq:field3}
\end{equation}%
For $A=1$
\begin{equation*}
\phi _{\beta }^{+}(x)_{1}=\int d\mu (p)(ue_{p}c_{\beta }(p)+ve_{-p}\tilde{c}%
_{\beta }(p)),
\end{equation*}%
and
\begin{equation*}
\phi _{\beta }^{-}(x)_{1}=\int d\mu (p)(ue_{-p}c_{\beta }(p)^{\dagger
}+ve_{p}\tilde{c}_{\beta }(p)^{\dagger }).
\end{equation*}%
Now the fields $\phi ^{+}$ and $\phi ^{-}$ describe the annihilation and
creation, respectively, parts of $\phi(x)_{A}$. In compact notation
\begin{eqnarray*}
\phi (x)_{J} &=&\phi _{\beta }^{+}(x)_{J}+\phi _{\beta }^{-}(x)_{J} \\
&=&\int d\mu (p)Q_{\beta }(x,p)_{JK}C_{\beta }(p)_{K} \\
&&+\int d\mu (p)\bar{Q}_{\beta }(x,p)_{JK}C_{\beta }(p)_{K}^{\dagger },
\end{eqnarray*}%
where $C_{\beta }(p)_{J}=(c_{\beta }(p),\tilde{c}_{\beta }(p))$,
\begin{equation}
Q_{\beta }(x,p)_{JK}=\left(
\begin{array}{cc}
ue_{p} & ve_{-p} \\
ve_{p} & ue_{-p}%
\end{array}%
\right).\label{eq:matrix1}
\end{equation}%
This notation for the fields will be useful in the demonstration of the Wick theorem. For the field at
zero temperature, the super-index $+(-)$ is associated with positive
(negative) frequencies. For the case of finite temperature, this notion is
lost, since there is a mixing of the frequencies. However, we still can use
this separation in components given in Eq.~(\ref{eq:field3}), as long as one
understands that it is not refereeing to the separation in positive-negative
frequencies.
The causal thermal propagator is defined by
\begin{eqnarray*}
\Delta (x-y;\beta )_{JK} &=&\langle 0;\beta |T~\phi (x)_{J}\phi
(y)_{K}|0;\beta \rangle \\
&=&\Theta (x_{0}-y_{0})\Delta _{>}(x,y;\beta )_{JK}+\Theta (y_{0}-x_{0})\Delta
_{<}(x,y;\beta )_{JK},
\end{eqnarray*}%
where $\Theta (x)$ is a Heaviside step function ($\Theta (x)=1$, if $x\geq 0$%
, and $0$ otherwise), $\Delta _{>}$ and $\Delta _{<}$ denote the advanced
and retarded propagators, respectively.

\section{Wick theorem in TFD}

In this section, the formalism of TFD for interacting fields is developed in
order to demonstrate the Wick theorem for thermal theories under algebraic
basis, achieving so in a very general result. We deal only with real scalar
field, although the generalization for charged field and spinor fields is
straightforward. In the following we set $\beta =1/T$ and chemical potential
$\mu =0$.

We now apply this scheme to interacting fields. The $n$-point correlation
function for an interacting field is defined as
\begin{equation}
\Delta (x_{1},...,x_{n};\beta )_{J_{1}~...~J_{n}}=\langle \Omega ;\beta |~T~\Phi
(x_{1})_{J_{1}}~...~\Phi (x_{n})_{J_{n}}|\Omega ;\beta \rangle ,
\label{eq:int-n-function1}
\end{equation}%
where $\Phi (x_{i})_{J_{i}}$ is an interacting field defined in Heisenberg
picture as
\begin{equation*}
\Phi (x_{i})_{J_{i}}=e^{it\hat{H}}~\phi (0,\vec{x}_{i})_{J_{i}}~e^{-it\hat{H}%
},
\end{equation*}%
with
\begin{equation*}
\hat{H}=H_{0}-\widetilde{H}_{0}+H_{\text{int}}-\widetilde{H}_{\text{int}%
}\equiv \hat{H}_{0}+\hat{H}_{\text{int}}
\end{equation*}%
being a duplicated Hamiltonian operator, and
\begin{equation*}
\Phi (x)_{J}=\Phi (x)_{J}^{+}+\Phi (x)_{J}^{-},
\end{equation*}%
so that
\begin{eqnarray*}
\Phi (x)_{1}^{+}=\int d\mu (p)~c(p)e^{ipx}, &~\hspace{0.3cm}~&\Phi
(x)_{2}^{-}=\int d\mu (p)~\widetilde{c}(p)e^{-ipx}, \\
\Phi (x)_{1}^{-}=\int d\mu (p)~c(p)^{\dagger }e^{-ipx}, &~\hspace{0.3cm}%
~&\Phi (x)_{2}^{-}=\int d\mu (p)~\widetilde{c}(p)^{\dagger }e^{+ipx}.
\end{eqnarray*}%
Using the matrix notation, introduced in Eq.s
(\ref{eq:field3}-\ref{eq:matrix1}), we have
\begin{equation*}
\Phi (x)_{J}=\int d\mu (p)~\left( Q_{\beta }(x,p)_{JK}C_{\beta }(p)_{K}+%
\bar{Q}_{\beta }(x,p)_{JK}C_{\beta }(p)_{K}\right) ,
\end{equation*}%

In the following we obtain the Gellmann-Low formula, defining a thermal
interacting vacuum. In interacting picture, let us introduce thermal fields
as
\begin{equation*}
\varphi (x_{i})_{J_{i}}=e^{it\hat{H}_{0}}~\phi (0,\vec{x}_{i})_{J_{i}}~e^{-it%
\hat{H}_{0}}.
\end{equation*}%
The thermal evolution operator from an instant $t_{1}$ to instant $t_{2}$ is
defined as
\begin{equation*}
\hat{U}(t_{1},t_{2})=U(t_{1},t_{2})~\widetilde{U}(t_{1},t_{2})=T~\exp \left(
-i\int_{t_{1}}^{t_{2}}dt~\hat{H}_{I}\right) ,
\end{equation*}%
such that
\begin{equation*}
\hat{H}_{I}=e^{it\hat{H}_{0}}~\hat{H}~e^{-it\hat{H}_{0}}.
\end{equation*}%
Recall that the Heisenberg and interaction pictures coincide with each other
at $t=0$.

The free thermal-field vacuum is defined as
\begin{equation*}
|0;\beta \rangle =Z_{0}(\beta )^{-1/2}~e^{-\beta \hat{H}_{0}}~|0\rangle ,~%
\hspace{0.3cm}~\text{with }\hspace*{0.3cm}Z_{0}(\beta )=\mathrm{Tr}%
(e^{-\beta \hat{H}_{0}}).
\end{equation*}%
Interacting thermal vacuum is defined similarly by
\begin{equation*}
|\Omega ;\beta \rangle =Z(\beta )^{-1/2}~e^{-\beta \hat{H}}~|0\rangle ,~%
\hspace{0.3cm}~\text{with }\hspace*{0.3cm}Z(\beta )=\mathrm{Tr}(e^{-\beta
\hat{H}}).
\end{equation*}%
Both vacua are related to one another by
\begin{eqnarray}
|\Omega ;\beta \rangle  &=&\left( \frac{Z_{0}(\beta )}{Z(\beta )}\right)
^{1/2}~\hat{U}(0,i\frac{\beta }{2})|0;\beta \rangle
\label{eq:int-thermal-vacuum1} \\
\langle \Omega ;\beta | &=&\langle \Omega ;\beta |\hat{U}(-i\frac{\beta }{2}%
,0)~\left( \frac{Z_{0}(\beta )}{Z(\beta )}\right) ^{1/2}.\label{eq:int-thermal-vacuum2}
\end{eqnarray}

We use these relations to obtain the Gellman-Low formula. Indeed, by using
Eqs.~(\ref{eq:int-thermal-vacuum1}) and (\ref{eq:int-thermal-vacuum2}) into
the expression of $n$-point function, Eq.~(\ref{eq:int-n-function1}), using
the definition of time-ordered operator together with some algebraic
manipulation and the definition of normalization factor $Z_{0}(\beta
)/Z(\beta )$, we thus obtain
\begin{equation}
\Delta (x_{1},...,x_{n};\beta )_{J_{1}~...~J_{n}}=\frac{\langle 0;\beta |~T~\hat{U}(-\infty
,\infty )~\varphi (x_{1})_{J_{1}}~...~\varphi (x_{n})_{J_{n}}|0;\beta
\rangle }{\langle 0;\beta |~T~\hat{U}(-\infty ,\infty )|0;\beta \rangle }.
\label{eq:Gellmann-Low}
\end{equation}%
Therefore, we describe the thermal $n$-point function for an
interacting
field theory in terms of free thermal fields, as long as we write $\hat{U}%
(-\infty ,\infty )$ in terms of free thermal fields. This happens for weak
coupling constants, which is proper for the development of perturbation
methods.

The Gellmann-low formula allows us to evaluate $n$-point functions
by perturbation methods. To derive such a formula, we expand the
expression
\begin{equation*}
\hat{U}(-\infty ,\infty )=T~\exp \left( -i\int_{-\infty }^{\infty }dt~\hat{H}%
_{I}(t)\right) =T~\exp \left( -i\int d^{4}x~\hat{\mathcal{H}}_{I}(x)\right) ,
\end{equation*}%
where $\hat{\mathcal{H}}_{I}(x)$ is a Hamiltonian density. In order to prove
the Wick theorem, it is usually assumed that $\hat{\mathcal{H}}_{I}(x)$ is
written as a sum of local products of fields $\varphi (x)_{J}$, e.g. $%
g_{n}\varphi (x)_{J}^{n}$. Then all we have to compute is a term like
\begin{equation*}
\langle 0;\beta |~T~\varphi (x_{1})_{J_{1}}~...~\varphi
(x_{n})_{J_{n}}|0;\beta \rangle .
\end{equation*}%
We hereafter perform this task for the case of $n=2$ in order to understand
the necessary steps. The generalization for an arbitrary $n$ will be
straightforwardly. The start point is then the expression
\begin{equation}
\langle 0;\beta |~T~\varphi (x_{1})_{J_{1}}\varphi (x_{2})_{J_{2}}|0;\beta
\rangle .  \label{eq:2-point1}
\end{equation}%
We observe in hindsight that the key step in this proof is to
consider a separation of fields in positive and negative
frequencies, considering a proper interpretation in case of
thermal fields.

Recalling that $\varphi (x_{i})_{J_{i}}$ is a free field at zero
temperature, the expectation value we want to compute is with respect to
thermal vacuum. Thus, we write these free fields in terms of thermal
creation and annihilation operators as
\begin{eqnarray*}
\phi (x_{i})_{J_{i}} &=&\phi (x_{i})_{J_{i}}^{+}+\phi
(x_{i})_{J_{i}}^{-} \\
&\equiv &\int d\mu (p)~\left( Q_{\beta }(x_{i},p)_{J_{i}K_{i}}C_{\beta
}(p)_{K_{i}}+\bar{Q}_{\beta }(x_{i},p)_{J_{i}K_{i}}C_{\beta }(p)_{K_{i}}\right) .
\end{eqnarray*}%
We stress once again that the above decomposition is not in terms of
positive and negative frequencies, once both frequencies mixed up inside each $%
\phi (x_{i})_{J_{i}}^{+}$ and $\phi (x_{i})_{J_{i}}^{-}$.

In order to compute the time-ordered product appearing inside 2-point
function given in Eq.~(\ref{eq:2-point1}), we introduce one more ingredient:
a normal ordering for thermal operators. This is defined by carrying all
thermal creation operators to the left of all thermal annihilation
operators. For instance,
\begin{equation*}
:c_{\beta }(p)\widetilde{c}_{\beta }(q)c_{\beta }(k)^{\dagger }:\equiv
c_{\beta }(k)^{\dagger }c_{\beta }(p)\widetilde{c}_{\beta }(q),
\end{equation*}%
where $:~:$ is used as the usual symbol for normal ordering
(although normal ordering is here being performed for thermal
creation and annihilation operators, and not the non-thermal
creation and annihilation operators). In addition, observe that
tilde operators commute with non-tilde operators. Using this
result, we compute the time-ordered product appearing inside
2-point function, Eq.~(\ref{eq:2-point1}). By simple algebraic
manipulations, one may bring the time-ordered product into a sum
of normal ordered products plus commutators of thermal fields.
Then we have
\begin{eqnarray*}
T~\phi (x_{1})_{J_{1}}\phi (x_{2})_{J_{2}} &=&~\theta
(x_{1}^{0}-x_{2}^{0})\left( :\phi (x_{1})_{J_{1}}\phi (x_{2})_{J_{2}}:~+~
\left[ \phi (x_{1})_{J_{1}}^{+},\phi (x_{2})_{J_{2}}^{-}\right] \right)  \\
&&+\theta (x_{2}^{0}-x_{1}^{0})\left( :\phi (x_{2})_{J_{2}}\phi
(x_{1})_{J_{1}}:~+~\left[ \phi (x_{2})_{J_{2}}^{+},\phi (x_{1})_{J_{1}}^{-}%
\right] \right) .
\end{eqnarray*}%
Since this time-ordered product is acting on thermal vacuum, thermal
normal-ordering product terms vanish. Therefore, we obtain
\begin{eqnarray*}
\langle 0;\beta |~T~\varphi (x_{1})_{J_{1}}\varphi (x_{2})_{J_{2}}|0;\beta
\rangle  &=&~\theta (x_{1}^{0}-x_{2}^{0})\left[ \phi
(x_{1})_{J_{1}}^{+},\phi (x_{2})_{J_{2}}^{-}\right]  \\
&&+~\theta (x_{2}^{0}-x_{1}^{0})\left[ \phi (x_{2})_{J_{2}}^{+},\phi
(x_{1})_{J_{1}}^{-}\right]  \\
&\equiv &~\Delta _{0}(x_{1}-x_{2};\beta )_{J_{1}J_{2}},
\end{eqnarray*}%
which is Feynman propagator for thermal free fields~\cite{Khanna09}.

The generalization to the $n$-point functions is straightforward. The
procedure is essentially the same we have done so far. Thus, the Wick
theorem for thermal scalar fields in general is written as
\begin{equation*}
\langle 0;\beta |~T~\varphi (x_{1})_{J_{1}}~...~\varphi
(x_{n})_{J_{n}}|0;\beta \rangle =\sum_{\pi }\prod_{i,j}^{n}\Delta
_{0}(x_{\pi (i)}-x_{\pi (j)};\beta )_{J_{\pi (i)}J_{\pi (i)}},
\end{equation*}%
i.e. the $n$-point function for thermal fields may be written as a product
of Feynman propagators for thermal free fields.

\section{Wick Theorem for Twisted Poincar\'e TFD}

In this section, we demonstrate the validity of the Wick theorem for the
twisted Poincar\'e quantum field theory at finite temperature using TFD
formalism. We begin with by recollecting some aspects of the twisted
Poincar\'e quantum field theory.

\subsection*{Twisted Quantization for Scalar Fields}

The main idea of twisted Poincar\'e quantum field theory is to
construct a Poincar\'e-invariant quantum field on the
Groenenwold-Moyal (GM) plane. The lesson is that we can perform
the usual quantization for one-particle states, i.e., obtaining
the usual unitary irreps of the Poincar\'e group. However, when we
deal with a many-particle system, we have to twist statistics of
fields. This is suitably seen as a twisting of the action of the
co-product of the Poincar\'e group. Notice that a co-product is
the algebraic structure that encodes the action of the underlying
symmetry group on tensor products. Therefore, it is the proper
structure to be considered when dealing with symmetries and
statistics of many-body systems.

Let $\mathcal{A}_{\theta }(\mathbb{R}^{D})$ be the algebra of smooth
functions on $\mathbb{R}^{D}$ with moyal product defined by the map
\begin{eqnarray*}
m_{\theta }:\mathcal{A}_{\theta }(\mathbb{R}^{D})\otimes \mathcal{A}_{\theta
}(\mathbb{R}^{D}) &\rightarrow &\mathcal{A}_{\theta }(\mathbb{R}^{D}) \\
f_{1}\otimes f_{2} &\mapsto &f_{1}\ast f_{2}\equiv m_{\theta }(f_{1}\otimes
f_{2}),
\end{eqnarray*}%
such that
\begin{equation*}
m_{\theta }(f_{1}\otimes f_{2})=m_{0}(F_{\theta }~f_{1}\otimes f_{2}),
\end{equation*}%
where $m_{0}$ is the point-wise product, i.e. $m_{0}(f_{1}\otimes
f_{2})\equiv f_{1}(x)\cdot f_{2}(x)$, and
\begin{equation*}
F_{\theta }=e^{\frac{i}{2}\theta ^{\mu \nu }\partial _{\mu }\otimes \partial
_{\nu }}
\end{equation*}%
is a \emph{twist element}, where $\theta ^{\mu \nu }=-\theta ^{\nu \mu }$ is
a constant matrix.

With this twisted product, we obtain the action of the co-product
of the Poincar\'e group $\mathcal{P}(\mathbb{R}^{D})$ by solving
the compatibility equation
\begin{equation}
m_{\theta }\left( \Delta _{\theta }(g)~f_{1}\otimes f_{2}\right)
=g~m_{\theta }(f_{1}\otimes _{2}),  \label{eq:compat-coprod-prod}
\end{equation}%
for any $f_{1},f_{2}\in \mathcal{A}_{\theta }(\mathbb{R}^{D})$,
with $g$ representing an element of the Poincar\'e group, and
$\Delta _{\theta }(g)$ being the action of twisted co-product of
$g$. In terms of the usual co-product $\Delta _{0}(g)=g\otimes g$,
the twisted co-product that solves the compatibility condition
given in Eq.~(\ref{eq:compat-coprod-prod}) is
\begin{equation*}
\Delta _{\theta }{g}=F_{\theta }\Delta _{\theta }{g}\tau _{0}\neq \tau
_{0}\Delta _{\theta }{g}~\hspace{0.3cm}~\text{for all }g\in \mathcal{P}(%
\mathbb{R}^{D}).
\end{equation*}

In order to study statistics, one usually considers a relation between the
flip operator
\begin{equation}
\tau _{0}(f_{1}\otimes f_{2})=f_{2}\otimes f_{2},  \label{eq:flip-op}
\end{equation}%
and the co-product of the isometry group. Since the co-product has been
twisted, it no longer commutes with the flip operator for all elements of
the Poincaré group, i.e.
\begin{equation*}
\Delta _{\theta }(g)\tau _{0}\neq \tau _{0}\Delta _{\theta }(g)~\hspace{0.3cm%
}~\text{for all }g\in \mathcal{P}(\mathbb{R}^{D}).
\end{equation*}%
However, if the flip operator is also twisted as
\begin{equation*}
\tau _{\theta }=F_{\theta }^{-1}~\tau _{0}~F_{\theta },
\end{equation*}%
then
\begin{equation*}
\Delta _{\theta }(g)\tau _{\theta }=\tau _{\theta }\Delta _{\theta }(g)~%
\hspace{0.3cm}~\text{for all }g\in \mathcal{P}(\mathbb{R}^{D}).
\end{equation*}%
But this twisting of the flip operator is equivalent to the twist of the
statistics of the fields, since we can define the following projectors
\begin{equation*}
P_{\theta }^{\pm }=\frac{\hat{1}\pm \tau _{\theta }}{2},
\end{equation*}%
where the super-index $+$ or $-$ here means projectors for boson or fermion
subspaces, respectively. In what follows, we consider bosons only.

A free scalar field invariant under the twisted Poincar\'e group
is written as
\begin{equation*}
\phi (x)=\int d\mu (p)~[a(p)e^{ipx}+a(p)^{\dagger }e^{-ipx}],
\end{equation*}%
where the twisted creation and annihilation operators $a(p)$ and $%
a(p)^{\dagger }$, respectively, satisfy
\begin{eqnarray*}
a(p)a(q) &=&~e^{i\theta ^{\mu \nu }p_{\mu }q_{\nu }}~a(q)a(p), \\
a(p)^{\dagger }a(q)^{\dagger } &=&~e^{i\theta ^{\mu \nu }p_{\mu }q_{\nu
}}~a(q)^{\dagger }a(p)^{\dagger }, \\
a(p)a(q)^{\dagger } &=&~e^{i\theta ^{\mu \nu }p_{\mu }q_{\nu
}}~a(p)^{\dagger }a(q)+2p^{0}\delta ^{(4)}(p-q).
\end{eqnarray*}%
An interacting Hamiltonian, in interaction representation, for these fields
may be written as
\begin{equation*}
H_{I}(t)=g_{n}\int d^{d}x~:\phi ^{\ast n}(x):=g_{n}\int d^{d}x~:\underbrace{%
\phi (x)\ast ...\ast \phi (x)}_{n}:,
\end{equation*}%
where
\begin{equation}
e^{ipx}\ast e^{iqx}=~e^{-\frac{i}{2}p_{\mu}\theta^{\mu\nu}q_{\nu}}~e^{i(p+q)x}.\label{astprod}
\end{equation}

In the following, we thermalize such a twisted theory exploring the
algebraic structure of TFD. This will represent an easy in the procedure, if
we attempt to perform the same calculation using the imaginary-time
approach, in particular to address the case of interacting fields.

\subsection*{Twisted Thermofield Dynamics}

In order to construct a twisted thermal field theory using TFD, we
observe that during such a development, the order that we use each
formalism does not matter. Indeed, one can first construct a TFD
for scalar field, by doubling of the Hilbert space and using the
Bogoliubov transformations, and then Poincar\'e twist on top of
resulting Hilbert space; or likewise, one can first Poincar\'e
twist the Hilbert space and then applies the TFD procedures on top
of the twisted Hilbert space.

Let us consider the doubled scalar field as given in Eqs.~(\ref{eq:field1})
and (\ref{eq:field2}). The generator of time translation is given by a
Hamiltonian in the form $\hat{H}=H-\tilde{H}$, such that
\begin{equation*}
\hat{H}=\int d\mu (p)~\omega (p)\left( c(p)c(p)^{\dagger }-\tilde{c}(p)%
\tilde{c}(p)^{\dagger }\right) .
\end{equation*}%
A total momentum operator is given by
\begin{equation*}
\hat{P}_{\nu }=P_{\nu }-\tilde{P}_{\nu }=\int d\mu (p)p_{\nu }(c(p)^{\dagger
}c(p)-\tilde{c}(p)^{\dagger }\tilde{c}(p)),
\end{equation*}%
with
\begin{eqnarray*}
\left[ \hat{P}_{\nu },c(p)\right]  &=&-p_{\nu }c(p),\qquad \left[ \hat{P}%
_{\nu },c(p)^{\dagger }\right] =p_{\nu }c(p)^{\dagger } \\
\left[ \hat{P}_{\nu },\tilde{c}(p)\right]  &=&-p_{\nu }\tilde{c}(p),\qquad %
\left[ \hat{P}_{\nu },\tilde{c}(p)^{\dagger }\right] =p_{\nu }\tilde{c}%
(p)^{\dagger }.
\end{eqnarray*}

We use $\hat{P}$ to construct a co-product, acting on $\hat{\mathcal{H}}$.
For that, we introduce the twisted annihilation and creation operators $%
a(p),a(p)^{\dagger }$ in terms of usual annihilation and creation operators $%
c(p),c(p)^{\dagger }$ as
\begin{eqnarray*}
a(p) &=&c(p)e^{\frac{i}{2}p_{\mu }\theta ^{\mu \nu }\hat{P}_{\nu }}, \\
a(p)^{\dagger } &=&e^{-\frac{i}{2}p_{\mu }\theta ^{\mu \nu }\hat{P}_{\nu
}}c(p)^{\dagger }, \\
\tilde{a}(p) &=&\tilde{c}(p)e^{\frac{i}{2}p_{\mu }\theta ^{\mu \nu }\hat{P}%
_{\nu }}, \\
\tilde{a}(p)^{\dagger } &=&e^{-\frac{i}{2}p_{\mu }\theta ^{\mu \nu }\hat{P}%
_{\nu }}\tilde{c}(p)^{\dagger },
\end{eqnarray*}%
where we have used $(\hat{P}_{\nu })\widetilde{}=-\hat{P_{\nu }}$. It is important to observe that, using the
twisted doubled operators, the momentum operator is written as
\begin{equation*}
\hat{P}_{\nu }=\int d\mu (p)~p_{\nu }~\left( a(p)^{\dagger }a(p)-\tilde{a}%
(p)^{\dagger }\tilde{a}(p)\right)
\end{equation*}

The Bogoliubov transformation is now defined by
\begin{equation*}
U_{\beta }=e^{-iG_{\beta}}
\end{equation*}%
with
\begin{equation*}
G_{\beta}=i\int d\mu (p)~f_{\beta }(p)~\left( a(p)^{\dagger }\tilde{%
a}(p)^{\dagger }-a(p)\tilde{a}(p)\right) ,
\end{equation*}%
where $f_{\beta }(p)$ is given in Eq.~(\ref{angle1}). The twisted thermal
fields are thus given by
\begin{eqnarray}
\phi _{\beta }(x) &=&\int d\mu (p)(a_{\beta }(p)e^{ipx}+a_{\beta
}(p)^{\dagger }e^{-ipx}),  \notag \\
\tilde{\phi}_{\beta }(x) &=&\int d\mu (p)(\tilde{a}_{\beta }(p)e^{-ipx}+%
\tilde{a}_{\beta }(p)^{\dagger }e^{ipx}),  \label{tphi}
\end{eqnarray}%
where the twisted thermal annihilation and creation operators are
\begin{eqnarray*}
a_{\beta }(p) &=&U_{\beta }~a(p)~U_{\beta }^{-1}, \\
a_{\beta }(p)^{\dagger } &=&U_{\beta }~a(p)^{\dagger }~U_{\beta }^{-1}.
\end{eqnarray*}%
In terms of untwisted operators, the above operators may be written as
\begin{eqnarray*}
a_{\beta }(p) &=&U_{\beta }~c(p)e^{\frac{i}{2}p_{\mu }\theta ^{\mu \nu }\hat{%
P}_{\nu }}~U_{\beta }^{-1}, \\
a_{\beta }(p)^{\dagger } &=&U_{\beta }~c(p)^{\dagger }e^{-\frac{i}{2}p_{\mu
}\theta ^{\mu \nu }\hat{P}_{\nu }}~U_{\beta }^{-1}.
\end{eqnarray*}%
Since
\begin{equation*}
\left[ \hat{P},G_{\beta }\right] =0,
\end{equation*}%
which can be easily checked by direct inspection, then we write
\begin{eqnarray}
a_{\beta }(p) &=&~c_{\beta }(p)e^{\frac{i}{2}p_{\mu }\theta ^{\mu \nu }\hat{P%
}_{\nu }},  \label{eq:tfdtwisted-ann-cre-3} \\
a_{\beta }(p)^{\dagger } &=&~e^{-\frac{i}{2}p_{\mu }\theta ^{\mu \nu }\hat{P}%
_{\nu }}c_{\beta }(p), \\
\tilde{a}_{\beta }(p) &=&~\tilde{c}_{\beta }(p)e^{\frac{i}{2}p_{\mu }\theta
^{\mu \nu }\hat{P}_{\nu }}, \\
\tilde{a}_{\beta }(p)^{\dagger } &=&~e^{-\frac{i}{2}p_{\mu }\theta ^{\mu \nu
}\hat{P}_{\nu }}\tilde{c}_{\beta }(p).
\end{eqnarray}

Let us now discuss $S$-matrices for interacting twisted thermal fields. We
first recall that in the usual TFD, the free-field 2-point function may be
written as
\begin{equation*}
\langle 0;\beta |T\phi (x)\phi (y)|0;\beta \rangle =\langle 0,0|T\phi
_{\beta }(x)\phi _{\beta }(y)|0,0\rangle .
\end{equation*}%
Then we may define the twisted thermal free-field 2-point function as
\begin{equation*}
\Delta _{0}^{\ast }(x-y;\beta )=\langle 0;\beta |T\phi (x)\ast \phi
(y)|0;\beta \rangle ,
\end{equation*}%
where the noncommutative product $\ast $ is defined in Eq.(\ref{astprod}).

In what follows we consider an interacting Hamiltonian in interaction
picture given in polynomial form as
\begin{eqnarray*}
\hat{H}_{I} &=&H_{I}-\tilde{H}_{I} \\
&=&\lambda \int d^{d}x~(:\phi ^{\ast n}(x):-:\tilde{\phi}^{\ast n}(x):),
\end{eqnarray*}%
where $:~:$ denotes normal ordering of the creation and annihilation
operators over the doubled Hilbert space\footnote{%
The normal ordering of the thermal operators is different because of the
mixing of the doubled operators due to Bogoliubov transformation}.

The $S$-matrix is defined as
\begin{eqnarray*}
\hat{S}_{\theta } &=&T\exp \left( -i\int dx^{0}\hat{H}_{I}{x^{0}}\right)  \\
&=&T\exp \left( -i\int dx^{d+1}(:\phi ^{\ast n}(x):-:\tilde{\phi}^{\ast
n}(x):)\right) .
\end{eqnarray*}

In \cite{Balachandran2006d}, it was proven that $\hat{S}_{\theta
}=\hat{S}_{0}$, in a
non-thermal formalism. This result is straightforwardly generalized to a
thermal formalism. Indeed, notice that twisted thermal fields can be written
in the form
\begin{equation*}
\phi _{\beta }(x)=\phi _{\beta 0}e^{\frac{i}{2}\overleftarrow{\partial }%
_{\mu }\theta ^{\mu \nu }\hat{P}_{\nu }},
\end{equation*}%
where $\phi _{\beta 0}$ is an untwisted thermal field constructed with
untwisted operators $c_{\beta }(p)$ and $c{\beta }(p)^{\dagger }$. Now, let
us consider the product of two plane wave vectors $e_{p}=e^{ipx}$ of
positive frequencies
\begin{equation*}
e_{p}\ast e_{q}=e^{-\frac{i}{2}p_{\mu }\theta ^{\mu \nu }q_{\nu }}e_{p+q}.
\end{equation*}%
Note that we consider $a(p)^{\dagger }=a(-p)$. Thus a twisted thermal field
may be written as
\begin{equation*}
\phi _{\beta }(x)=\int d\mu (p)~\left( a_{\beta }(p)e_{p}+a_{\beta
}(-p)e_{-p}\right) .
\end{equation*}%
For $n=2$ in $\hat{H}$, the first term of the $\hat{S}_{\theta }$ expansion
is 
\begin{equation*}
\hat{S}_{\theta }^{(1)}=-i\lambda \int dx^{d+1}(:\phi \ast \phi :-:\tilde{%
\phi}\ast \tilde{\phi}:).
\end{equation*}%
As a consequence, we have terms of the type
\begin{equation*}
a_{\beta }(p)a_{\beta }(q)e_{p}\ast e_{q}=a_{\beta }(p)a_{\beta }(q)e^{-%
\frac{i}{2}p_{\mu }\theta ^{\mu \nu }q_{\nu }}e_{p+q}.
\end{equation*}%
With the expression, we derive 
\begin{eqnarray}
a_{\beta }(p)a_{\beta }(q)e_{p}\ast e_{q} &=&c_{\beta }(p)e^{\frac{i}{2}%
p_{\mu }\theta ^{\mu \nu }\hat{P}_{\nu }}c_{\beta }(q)e^{\frac{i}{2}q_{\mu
}\theta ^{\mu \nu }\hat{P}_{\nu }}e^{-\frac{i}{2}p_{\mu }\theta ^{\mu \nu
}q^{\nu }}e_{p+q} \\
&=&c(p)c(q)e_{p+q}e^{\frac{i}{2}(p+q)_{\mu }\theta
^{\mu \nu }\hat{P}\nu }.  \label{cinco}
\end{eqnarray}%
Writing $\partial _{\mu }e_{p+q}=i(p+q)_{\mu }e_{p+q},$ we have
\begin{equation*}
e_{p+q}e^{\frac{i}{2}(p+q)_{\mu }\theta ^{\mu \nu }\hat{P}_{\nu }}=e_{p+q}e^{%
\frac{i}{2}\overleftarrow{\partial }_{\mu }\theta ^{\mu \nu }\hat{P}_{\nu }},
\end{equation*}%
and
\begin{equation*}
a_{\beta }(p)a_{\beta }(q)e_{p}\ast e_{q}=a_{\beta }(p)a_{\beta }(q)e^{\frac{%
i}{2}\overleftarrow{\partial }_{\mu }\theta ^{\mu \nu }\hat{P}_{\nu }}.
\end{equation*}%
Thus the $S$-matrix term of first order in $\lambda $ is
\begin{equation*}
\hat{S}_{\theta }^{(1)}=-i\lambda \int d^{d+1}x(:\phi ^{2}:-:\tilde{\phi}%
^{2}:)e^{\frac{i}{2}\overleftarrow{\partial }_{\mu }\theta ^{\mu \nu }\hat{P}%
_{\nu }}.
\end{equation*}%
Considering that the interaction does not involve long range forces, we
expand the exponential in the integrand and discard surface terms. We
finally obtain
\begin{equation*}
\hat{S}_{\theta }^{(1)}=\hat{S}_{0}^{(1)}.
\end{equation*}

The same procedure goes through in second order in $\lambda $
\begin{eqnarray*}
\hat{S}_{\theta }^{(2)}&=&\frac{(-i\lambda ^{2})}{2!}\int
d^{d+1}x_{1}d^{d+1}x_{2}  [\Theta (x_{1}^{0}-x_{2}^{0})(:(\phi
\ast \phi
)(x_{1}):\ast :(\phi \ast \phi )(x_{2}): \\
&&+\Theta (x_{2}^{0}-x_{1}^{0})( :(\phi \ast \phi )(x_{2}):\ast
:(\phi \ast
\phi )(x_{1}):) \\
&&-\Theta (x_{1}^{0}-x_{2}^{0})) :(\tilde{\phi}\ast \tilde{\phi}%
)(x_{1}):\ast :(\tilde{\phi}\ast \tilde{\phi})(x_{2}):) \\
&&-\Theta (x_{2}^{0}-x_{1}^{0})( :(\tilde{\phi}\ast \tilde{\phi}%
)(x_{2}):\ast :(\tilde{\phi}\ast \tilde{\phi})(x_{1}):)].
\end{eqnarray*}%
In terms of twisted thermal annihilation and creation operators, the above
expression displays terms into the form
\begin{eqnarray*}
T_{\beta }^{\ast } &=&\Theta (x_{1}^{0}-x_{2}^{0}):a_{\beta }(p_{1})a_{\beta
}(q_{1})::a_{\beta }(p_{2})a_{\beta }(q_{2}): \\
&&\times (e_{p_{1}}\ast e_{q_{1}})(x_{1})\ast (e_{p_{2}}\ast
e_{q_{2}})(x_{2}).
\end{eqnarray*}%
Using Eq. (\ref{cinco}), we have
\begin{eqnarray}
T_{\beta }^{\ast } &=&\Theta (x_{1}^{0}-x_{2}^{0})\left[ :c_{\dagger
}(p_{1})c_{\dagger }::c_{\dagger }(p_{2})c_{\dagger }:e^{-\frac{i}{2}%
(p_{1}+q_{1})\mu \theta ^{\mu \nu }(p_{2}+q_{2}\nu )}\right.  \notag \\
&&\times \left. \left( e_{p_{1}+q_{1}}(x_{1})e_{p_{2}+q_{2}}(x_{2})e^{\frac{1%
}{2}(\frac{\overleftarrow{\partial }}{\partial x_{1\mu }}+\frac{%
\overleftarrow{\partial }}{\partial x_{2\mu }})\theta ^{\mu \nu }\hat{P}%
_{\nu }}\right) \right] .  \label{tres}
\end{eqnarray}%
The momentum conservation is expressed as
\begin{equation*}
e^{-\frac{i}{2}(p_{1}+q_{1})\mu \theta ^{\mu \nu }(p_{2}+q_{2})_{\nu }}=1.
\end{equation*}%
Then we show that the $S$-matrix, to all orders of $\lambda ,$ satisfies $%
\hat{S}_{\theta }^{(2)}=\hat{S}_{0}^{(2)}.$ Therefore, for $n=2$ we have
\begin{equation*}
\hat{S}_{\theta }=\hat{S}_{0}.
\end{equation*}

Let us now obtain a Wick theorem for the twisted TFD. First, we calculate
expressions like $\langle 0|T\phi _{\beta J_{1}}\ast \phi _{\beta J_{2}}\ast
\cdots \ast \phi _{\beta J_{n}}|0\rangle ,$ or, equivalently
\begin{equation*}
\langle 0;\beta |T\phi _{J_{1}}\ast \phi _{J_{2}}\ast \cdots \ast \phi
_{J_{n}}|0;\beta \rangle .
\end{equation*}%
where $\phi _{J}(x)=(\phi (x),\tilde{\phi}(x)^{\dagger })$. The Fourier
representation for these twisted doubled fields is
\begin{eqnarray*}
\phi (x)_{J} &=&\int d\mu (p)~\left( a(p)_{J}e_{p}+a(p)_{J}^{\dagger
}e_{-p}\right) \\
&=&\phi ^{+}(x)_{J}+\phi ^{-}(x)_{J},
\end{eqnarray*}%
and
\begin{eqnarray*}
\phi ^{+}(x)_{1} &=&\int d\mu (p)a(p)e_{p},\qquad \phi ^{-}(x)_{2}=\int d\mu
(p)\tilde{a}(p)e_{-p} \\
\phi ^{-}(x)_{1} &=&\int d\mu (p)a^{\dagger }(p)e_{-p},\quad \phi
^{+}(x)_{2}=\int d\mu (p)\tilde{a}(p)^{\dagger }e_{p}.
\end{eqnarray*}

We write now the twisted operators in terms of thermal operators. For that
we use the Bogoliubov transformation, such that
\begin{equation*}
a(p)_{J}=(U_{\beta }^{-1})_{JK}~a_{\beta }(p)_{K},
\end{equation*}%
with
\begin{equation*}
a(p)_{J}=\left(
\begin{array}{c}
a(p) \\
\tilde{a}(p)^{\dagger }%
\end{array}%
\right) \qquad a_{\beta}(p)_{J}=\left(
\begin{array}{c}
a_{\beta }(p) \\
\tilde{a}_{\beta }(p)^{\dagger }%
\end{array}%
\right) .
\end{equation*}%
Then we write $\phi _{\beta }(x)_{J}$, the representation of the doubled
field in terms of thermal operators, as
\begin{equation*}
\phi (x)_{J}=\phi _{\beta }^{+}(x)_{J}+\phi _{\beta }^{-}(x)_{J}.
\end{equation*}%
For $J=1$
\begin{equation*}
\phi _{\beta }^{+}(x)_{1}=\int d\mu (p)(ue_{p}a_{\beta }(p)+ve_{-p}\tilde{a}%
_{\beta }(p)),
\end{equation*}%
and
\begin{equation*}
\phi _{\beta }^{-}(x)_{1}=\int d\mu (p)(ue_{-p}a_{\beta }(p)^{\dagger
}+ve_{p}\tilde{a}_{\beta }(p)^{\dagger }).
\end{equation*}%
Now the fields $\phi ^{+}$ and $\phi ^{-}$ describe the annihilation and
creation, respectively, parts of $\phi(x)$. In compact notation
\begin{eqnarray*}
\phi (x)_{J} &=&\phi _{\beta }^{+}(x)_{J}+\phi _{\beta }^{-}(x)_{J} \\
&=&\int d\mu (p)Q_{\beta }(x,p)_{JK}A_{\beta }(p)_{K} \\
&&+\int d\mu (p)\bar{Q}_{\beta }(x,p)_{JK}A_{\beta }(p)_{K}^{\dagger },
\end{eqnarray*}%
where $A_{\beta }(p)_{J}=(a_{\beta }(p),\tilde{a}_{\beta }(p))$, and
\begin{equation*}
Q_{\beta }(x,p)_{JK}=\left(
\begin{array}{cc}
ue_{p} & ve_{-p} \\
ve_{p} & ue_{-p}%
\end{array}%
\right)
\end{equation*}%
Writing the twisted operators in terms of the untwisted creation and annihilation operators we obtain
\begin{equation*}
A_{\beta }(p)_{J}=\left(
\begin{array}{c}
c_{\beta }(p)e^{\frac{i}{2}p_{\mu }\theta ^{\mu \nu }\hat{P}_{\nu }} \\
\tilde{c}_{\beta }(p)e^{\frac{i}{2}p_{\mu }\theta ^{\mu \nu }\hat{P}_{\nu }}%
\end{array}%
\right) =\left(
\begin{array}{c}
c_{\beta }(p) \\
\tilde{c}_{\beta }(p)%
\end{array}%
\right) e^{\frac{i}{2}p_{\mu }\theta ^{\mu \nu }\hat{P}_{\nu }}
\end{equation*}%
and
\begin{equation*}
A_{\beta }(p)_{J}^{\dagger }=\left(
\begin{array}{c}
c_{\beta }(p)^{\dagger } \\
\tilde{c}_{\beta }(p)^{\dagger }%
\end{array}%
\right) e^{-\frac{i}{2}p_{\mu }\theta ^{\mu \nu }\hat{P}_{\nu }},
\end{equation*}%
we have the compact notation
\begin{eqnarray}
A_{\beta }(p)_{J} &=&C_{\beta }(p)_{J}e^{\frac{i}{2}p_{\mu }\theta ^{\mu \nu
}\hat{P}_{\nu }}, \\
A_{\beta }(p)_{J}^{\dagger } &=&C_{\beta }(p)_{J}^{\dagger }e^{-\frac{i}{2}%
p_{\mu }\theta ^{\mu \nu }\hat{P}_{\nu }}.  \label{seis}
\end{eqnarray}%
Now we are able to calculate the products $\phi _{\beta }^{\pm
}(x)_{J}\ast \phi _{\beta }^{\pm }(y)_{K}$ in
\begin{eqnarray*}
\phi _{\beta }(x)_{J}\ast \phi _{\beta }(y)_{K} &=&\int d\mu (p)d\mu (q)[Q_{\beta
}(x,p)_{JJ^{\prime }}A_{\beta }(p)_{J^{\prime }}Q_{\beta }(y,q)_{KK^{\prime
}}A_{\beta }(q)_{K^{\prime }} \\
&&+Q_{\beta }(x,p)_{JJ^{\prime }}A_{\beta }(p)_{J^{\prime }}\bar{Q}%
_{\beta }(y,q)_{KK^{\prime }}A_{\beta }(q)_{K^{\prime }}^{\dagger } \\
&&+\bar{Q}_{\beta }(x,p)_{JJ^{\prime }}A_{\beta
}(p)_{J^{\prime
}}^{\dagger }Q_{\beta }(y,q)_{KK^{\prime }}A_{\beta }(q)_{K^{\prime }} \\
&&+\bar{Q}_{\beta }(x,p)_{JJ^{\prime }}A_{\beta
}(p)_{J^{\prime }}^{\dagger }\bar{Q}_{\beta
}(y,q)_{KK^{\prime }}A_{\beta }(q)_{K^{\prime }}^{\dagger }].
\end{eqnarray*}%
The first term involves
\begin{eqnarray*}
\phi _{\beta }^{+}(x)_{1}\ast \phi _{\beta }^{+}(y)_{1} &=&A_{\beta }(p)_{1}A_{\beta
}(q)_{1}Q_{\beta }(x,p)_{11}\ast Q_{\beta }(y,q)_{11} \\
&&+A_{\beta }(p)_{1}A_{\beta }(q)_{2}Q_{\beta
}(x,p)_{11}\ast Q_{\beta }(y,q)_{12} \\
&&+A_{\beta }(p)_{2}A_{\beta }(q)_{1}Q_{\beta }(x,p)_{12}\ast Q_{\beta
}(y,q)_{11} \\
&&+A_{\beta }(p)_{2}A_{\beta }(q)_{2}Q_{\beta }(x,p)_{12}\ast Q_{\beta
}(y,q)_{12}
\end{eqnarray*}%
Notice that
\begin{equation*}
Q_{\beta }(x,p)_{JK}=U_{JK}^{-1}E(x,p)_{JK}
\end{equation*}%
with
\begin{equation*}
E(x,p)_{JK}=\left(
\begin{array}{cc}
e_{p} & e_{-p} \\
e_{p} & e_{-p}%
\end{array}%
\right) .
\end{equation*}%
Then the products of the Q-matrix are given by
\begin{equation*}
Q_{\beta }(x,p)_{JK}\ast Q_{\beta }(y,q)_{LM}=Q_{\beta }(x,p)_{JK}Q_{\beta
}(y,q)_{LM}e^{\frac{i}{2}\overline{p}_{\mu }\theta ^{\mu \nu }\overline{q}%
_{\nu }},
\end{equation*}%
with
\begin{equation*}
\overline{p}_{\mu }=%
\begin{cases}
+p_{\mu } & \text{if $j=1$}; \\
-p_{\mu } & \text{if $j=2$}.%
\end{cases}%
\end{equation*}%
Using Eqs.~(\ref{cinco}) and (\ref{seis}), we obtain
\begin{equation}
\phi _{\beta }^{\pm }(x)_{J}\ast \phi _{\beta }^{\pm }(y)_{K}=\phi _{\beta }^{\pm
}(x)_{J}\phi _{\beta }^{\pm }(y)_{K}e^{\frac{i}{2}(\overline{p}+\overline{q})_{\mu
}\theta ^{\mu \nu }\hat{P}_{\nu }}.  \label{sete}
\end{equation}%
With the twisted co-product, given in Eq.~(\ref{sete}), we write
the time ordering in terms of the normal ordering
\begin{eqnarray*}
\mathcal{D}_{JK}(x_{1}-x_{2}) &=&T\phi (x_{1})_{J}\ast \phi (x_{2})_{K} \\
&=&\theta (x_{1}^{0}-x_{2}^{0})D_{JK}(x_{1}-x_{2})+\theta
(x_{2}^{0}-x_{1}^{0})D_{KJ}(x_{1}-x_{2})
\end{eqnarray*}%
where%
\begin{equation*}
D_{JK}(x_{1}-x_{2})=(\phi _{\beta }^{+}(x_{1})_{J}+\phi _{\beta
}^{-}(x_{1})_{J})\ast (\phi _{\beta }^{+}(x_{2})_{K}+\phi _{\beta
}^{-}(x_{2})_{K})
\end{equation*}%
and%
\begin{equation*}
D_{KJ}(x_{1}-x_{2})=\theta (x_{2}^{0}-x_{1}^{0})(\phi _{\beta
}^{+}(x_{2})_{K}+\phi _{\beta }^{-}(x_{2})_{K})\ast (\phi _{\beta
}^{+}(x_{1})_{J}+\phi _{\beta }^{-}(x_{1})_{J}).
\end{equation*}%
We write $D_{JK}(x_{1}-x_{2})$ as
\begin{eqnarray*}
D_{JK}(x_{1}-x_{2}) &=&(\phi _{\beta }^{+}(x_{1})_{J}+\phi _{\beta
}^{-}(x_{1})_{J})\ast (\phi _{\beta }^{+}(x_{2})_{K}+\phi _{\beta
}^{-}(x_{2})_{K}) \\
&=&\underline{\phi _{\beta }^{+}(x_{1})_{J}\ast \phi _{\beta }^{+}(x_{2})_{K}%
}+\phi _{\beta }^{+}(x_{1})_{J}\ast \phi _{\beta }^{-}(x_{2})_{K} \\
&&+\underline{\phi _{\beta }^{-}(x_{1})_{J}\ast \phi _{\beta }^{+}(x_{2})_{K}%
}+\underline{\phi _{\beta }^{-}(x_{1})_{J}\ast \phi _{\beta }^{1}(x_{2})_{K}}%
,
\end{eqnarray*}%
where the underlined terms are normal ordered with respect to the thermal
operators. Using the deformed product commutator $\left[ \phi _{\beta
}^{+}(x_{1})_{J},\phi _{\beta }^{-}(x_{2})_{K}\right] ^{\ast }$, we have
\begin{eqnarray*}
D_{JK}(x_{1}-x_{2}) &=&\underline{\phi _{\beta }^{+}(x_{1})_{J}\ast \phi
_{\beta }^{+}(x_{2})_{K}}+\underline{\phi _{\beta }^{-}(x_{2})_{K}\ast \phi
_{\beta }^{+}(x_{1})_{J}} \\
&&+\underline{\phi _{\beta }^{-}(x_{1})_{J}\ast \phi _{\beta }^{+}(x_{2})_{K}%
}+\underline{\phi _{\beta }^{-}(x_{1})_{J}\ast \phi _{\beta }^{-}(x_{2})_{K}}
\\
&&+\left[ \phi _{\beta }^{+}(x_{1})_{J},\phi _{\beta }^{-}(x_{2})_{K}\right]
^{\ast }.
\end{eqnarray*}%
Relatively to thermal operators the normal ordering is denoted as
\begin{eqnarray*}
N_{\theta }a_{\beta }(p)a_{\beta }(q)^{\dagger } &=&a_{\beta }(q)^{\dagger
}a_{\beta }(p) \\
N_{\theta }a_{\beta }(p)^{\dagger }a_{\beta }(q) &=&a_{\beta }(p)^{\dagger
}a_{\beta }(p) \\
N_{\theta }a_{\beta }(p_{1})\tilde{a}_{\beta }(q_{1})^{\dagger }\tilde{a}%
_{\beta }(p_{2})a_{\beta }(q_{2})^{\dagger } &=&\tilde{a}_{\beta
}(q_{1})^{\dagger }a_{\beta }(q_{2})^{\dagger }\tilde{a}_{\beta
}(p_{2})a_{\beta }(p_{1}) \\
&\vdots &
\end{eqnarray*}%
Thus we obtain
\begin{eqnarray*}
T\phi (x_{1})_{J}\ast \phi (x_{2})_{K} &=&N_{\theta }\phi (x_{1})_{J}\ast
\phi (x_{2})_{K} \\
&&+\theta (x_{1}^{0}-x_{2}^{0})\left[ \phi _{\beta }^{+}(x_{1})_{J},\phi
_{\beta }^{-}(x_{2})_{K}\right] ^{\ast } \\
&&+\theta (x_{2}^{0}-x_{1}^{0})\left[ \phi _{\beta }^{+}(x_{2})_{K},\phi
_{\beta }^{-}(x_{1})_{J}\right] ^{\ast }.
\end{eqnarray*}%
Defining the contraction of the fields as
\begin{eqnarray*}
\contraction{}{\phi}{(x_1)_{J}\ast}{{\phi}}\phi (x_{1})_{J}\ast{\phi }(x_{2})_{K}
&=&+\theta (x_{1}^{0}-x_{2}^{0})\left[ \phi _{\beta }^{+}(x_{1})_{J},\phi
_{\beta }^{-}(x_{2})_{K}\right] ^{\ast } \\
&&+\theta (x_{2}^{0}-x_{1}^{0})\left[ \phi _{\beta }^{+}(x_{2})_{K},\phi
_{\beta }^{-}(x_{1})_{J}\right] ^{\ast },
\end{eqnarray*}%
we obtain
\begin{equation*}
T\phi(x_1)_{J}\ast\phi(x_2)_{K}=N_\theta\phi(x_1)_{J}%
\ast\phi(x_2)_{K}+\contraction{}{\phi}{(x_1)_{J}\ast}{{\phi}}\phi (x_{1})_{J}\ast{\phi }(x_{2})_{K}
\end{equation*}%
Therefore
\begin{eqnarray*}
\langle 0;\beta |T\phi (x_{1})_{J}\ast \phi (x_{2})_{K}|0;\beta \rangle
&=&\langle 0;\beta |N_{\theta }\phi (x_{1})_{J}\ast \phi (x_{2})_{K}|0;\beta
\rangle  \\
&&+\langle 0;\beta |\contraction{}{\phi}{(x_1)_{J}\ast}{{\phi}}\phi (x_{1})_{J}\ast{%
\phi }(x_{2})_{K}|0;\beta \rangle
\end{eqnarray*}%
The last expression is our guide for defining the Feynman propagator for the
twisted TFD, i.e.
\begin{eqnarray*}
\Delta _{0}^{\ast }(x_{1}-x_{2};\beta )_{JK} &=&\langle 0;\beta |%
\contraction{}{\phi}{(x_1)_{J}\ast}{{\phi}}\phi (x_{1})_{J}\ast {\phi }%
(x_{2})_{K}|0;\beta \rangle  \\
&=&\langle 0;\beta |\contraction{}{\phi}{(x_1)_{J}}{{\phi}}\phi (x_{1})_{J}{%
\phi }(x_{2})_{K}e^{\frac{i}{2}(\overline{p}+\overline{q})_{\mu }\theta
^{\mu \nu }\hat{P}_{\nu }}|0;\beta \rangle  \\
&=& \Delta _{0}(x_{1}-x_{2};\beta )_{JK}
\end{eqnarray*}%
where we expanded the exponential and used the fact that $\hat{P}_{\mu
}|0;\beta \rangle =0$.

For any number of fields we are interesting in products of the type%
\begin{eqnarray*}
\Pi _{i}^{\ast }\phi (x_{1})_{J_{i}} &=&\phi (x_{1})_{J_{1}}\ast \phi
(x_{2})_{J_{2}}\ast \cdots \ast \phi (x_{n})_{J_{n}} \\
&=&(\phi _{\beta }^{+}(x_{1})_{J_{1}}+\phi _{\beta }^{-}(x_{1})_{J_{1}})\ast
\cdots \ast (\phi _{\beta }^{+}(x_{n})_{J_{n}}+\phi _{\beta
}^{-}(x_{n})_{J_{n}}).
\end{eqnarray*}%
We have then $2^{n}$ products of the form
\begin{equation}
\Pi _{i}^{\pm \ast }\phi (x_{i})_{J_{i}}=\phi _{\beta }^{\pm
}(x_{1})_{J_{1}}\ast \cdots \ast \phi _{\beta }^{\pm }(x_{n})_{J_{n}}.
\label{nproduct}
\end{equation}%
With each of the fields in terms of untwisted operators we have
\begin{equation*}
\Pi _{i}^{\pm \ast }\phi (x_{i})_{J_{i}}=\phi _{\beta }^{\pm
}(x_{1})_{J_{1}}\cdots \phi _{\beta }^{\pm }(x_{n})_{J_{n}}e^{\frac{i}{2}%
\sum_{j=1}^{n}(p_{j})\mu \theta ^{\mu \nu }\hat{P}_{\nu }}.
\end{equation*}%
This implies, together with$\hat{P}_{\mu }|0;\beta \rangle =0$, that
\begin{equation*}
\langle 0;\beta |T\phi (x_{1})_{J_{1}}\ast \cdots \ast \phi
(x_{n})_{J_{n}}|0;\beta \rangle =\langle 0;\beta |T\phi
(x_{1})_{J_{1}}\cdots \phi (x_{n})_{J_{n}}|0;\beta \rangle .
\end{equation*}%
Therefore, the Wick theorem for twisted TFD is
\begin{equation*}
T\prod_{i=1}^{n}\phi (x_{i})_{J_{i}}=\sum_{k=0}^{\left\lfloor \frac{n}{2}%
\right\rfloor }\sum_{\{\pi \}}N_{\theta }\prod_{j=1}^{n-2k}\phi (x_{\pi
j})_{J_{\pi j}}\prod_{\{(\pi l^{\prime },\pi l)\}}G_{0}(x_{\pi l^{\prime
}}-x_{\pi l};\beta )_{J_{\pi l^{\prime }}J_{\pi l}}
\end{equation*}%
where $\pi =(\pi _{1},\cdots ,\pi _{n})$ is a permutation $(1,\cdots ,n)$,
with $\pi _{j}$ an element of the combination $(1,...,n)$ token $n-2k$ a $%
n-2k$, for $j=1,...,n-2k$. $\{(\pi l,\pi l^{\prime })\}$ is the set of all
pairs inside $\pi l^{\prime },\pi l$ for $n-2k\leq l,l^{\prime }\leq n$,
with $l\neq l^{\prime }$. This result shows that the TFD Wick theorem
remains unchanged.

\section{Concluding remarks}

In this work we have developed a perturbative approach for twisted
Poincar\'e invariant theories at finite temperature. We
have used the algebraic formalism thermofield dynamics in order to
derive the thermal Wick theorem for this type of non-commutative
theory. During the demonstration, we rederive the Wick theorem for
TFD, but under very general basis, a result that is not available
in the literature, at least at our best knowledge.

Our results can be readily extended to other fields, as fermion and gauge
fields. This aspect will be addressed in detail in another place.

\bigskip

\section*{Acknowledgements}

The authors thank  CNPq and CAPES (of Brazil) for financial
support. ARQ acknowledges  CNPq for support under grant no.
307760/2009-0.

%\bibliography{referencias}
%\bibliographystyle{JHEP}

\providecommand{\href}[2]{#2}\begingroup\raggedright\endgroup

\end{document}